\documentclass[fleqn,usenatbib]{mnras}

\usepackage{newtxtext,newtxmath}
\usepackage[T1]{fontenc}
\usepackage{ae,aecompl}

\usepackage{graphicx}	% Including figure files
\usepackage{amsmath}	% Advanced maths commands
\usepackage{amssymb}	% Extra maths symbols

\title[Newtonian cosmologies with time-varying $G$]{Low-redshift tests of Newtonian
cosmologies with a time-varying gravitational constant}

\author[E.T. Han{\i}meli et al.]{
Ekim Taylan Han{\i}meli,$^{1,2}\thanks{E-mail: ekim.hanimeli@zarm.uni-bremen.de}\thanks{Current address: ZARM, Universit{\"a}t Bremen, Am Fallturm, D-28359 Bremen, Germany}$
Isaac Tutusaus,$^{3,4,1}$ 
Brahim Lamine,$^{1}$ 
Alain Blanchard$^{1}$
\\
$^{1}$Universit{\'e} de Toulouse, UPS-OMP, IRAP, CNRS, 14 Avenue Edouard Belin, F-31400 Toulouse, France\\
$^{2}$Lule{\aa} University of Technology, Space Campus, Rymdcampus 1, 981 92 Kiruna, SE-Sweden\\
$^{3}$Institute of Space Sciences (ICE, CSIC), Campus UAB, Carrer de Can Magrans, s/n, E-08193 Barcelona, Spain\\
$^{4}$Institut d'Estudis Espacials de Catalunya (IEEC), Carrer Gran Capit\`a 2-4, E-08193 Barcelona, Spain
}

\date{Accepted XXX. Received YYY; in original form ZZZ}

\pubyear{2020}

\usepackage{xcolor}
\def\dd{\mathrm{d}}

\begin{document}
\label{firstpage}
\pagerange{\pageref{firstpage}--\pageref{lastpage}}
\maketitle

\begin{abstract}
In this work, we investigate Newtonian cosmologies with a time-varying gravitational constant, $G(t)$. 
We examine  whether such models can reproduce the low-redshift cosmological observations without a cosmological constant, or any other sort of explicit dark energy fluid.
Starting with a modified Newton's second law, where $G$ is taken as a function of time, we derive the first Friedmann--Lema{\^i}tre equation, where a second parameter, $G^*$, appears as the gravitational constant. This parameter is related to the original $G$ from the second law, which remains in the acceleration equation.
We use this approach to reproduce various cosmological scenarios that are studied in the literature, and we test these models with low-redshift probes: type-Ia supernovae (SNIa), baryon acoustic oscillations, and cosmic chronometers, taking also into account a possible change in the supernovae intrinsic luminosity with redshift. 
As a result, we obtain several models
with similar $\chi^2$ values as the standard $\Lambda$CDM cosmology. When we allow for a redshift-dependence of the SNIa intrinsic luminosity, 
a model with a $G$ exponentially decreasing to zero while remaining positive (model 4) can explain the observations without acceleration. When we assume  no redshift-dependence of SNIa, 
the observations favour a negative $G$
at large scales, while $G^*$ remains positive for most of these models.
We conclude that these models offer interesting interpretations to the low-redshift cosmological observations, without needing a dark energy term.

\end{abstract}


%%%%%%%%%%%%%%%%%%%%%%%%%%%%%%%%%%%%%%%%%%%%%%%%%%

%%%%%%%%%%%%%%%%% BODY OF PAPER %%%%%%%%%%%%%%%%%%

\section{Introduction}
\label{sec:intro}

The accelerating expansion of the Universe~\citep{Riess1998,Perlmutter1999} is a widely accepted idea with a large amount of observational support (see e.g. the reviews~\citet{Blanchard2010,obsprobes} and references therein). In the standard model of cosmology, $\Lambda$CDM, this acceleration is interpreted as a cosmological constant, $\Lambda$, which acts as the vacuum energy with positive energy density but negative pressure. However, the value of this constant is at odds with quantum field theoretical estimations of the vacuum energy, which creates the cosmological constant problem~\citep{weinberg}. Moreover, on the observational side, there has recently been a debate in the literature about whether type-Ia supernovae (SNIa) data (either alone or combined with other probes) can definitely prove the accelerated nature of the expansion of the Universe~\citep{Sarkar,Ringermacher,Rubin,Shariff,Dam,Haridasu,isaac2017,Lin,Lonappan,Lukovi,Colin2018,isaac2018}. Therefore, it is reasonable to question this standard cosmological picture from both theoretical and observational perspectives.

One widely considered alternative is the possibility of the gravitational constant changing over cosmological time.
This is in fact an old idea, dating back at least to \citet{dirac} and his large numbers hypothesis.
Accordingly, in the literature there are multiple ways of implementing such a variation.
One popular way is to replace the gravitational constant with a dynamical scalar field evolving with time (and space).
The most extensively studied relativistic models among these are the so-called Jordan--Brans--Dicke (JBD) theories~\citep{jordan,bransdicke}. These theories, and some of their extensions, such as scalar--tensor theories, have been confronted with observations for many decades, and very tight constraints have been put on them with solar system measurements. In particular, the original JBD model introduces a dimensionless parameter $\omega$, which should naturally be of the order of unity, but is constrained to be larger than roughly $40\,000$ by solar system tests~\citep{bertotti}.
In order to overcome these solar system constraints, screening mechanisms can be added to these theories to separate their predictions in cosmological scales from the local observations (for a more detailed discussion of various screening mechanisms see~\citet{joyce}).
Besides these JBD variations, several other approaches for obtaining a variable gravitational constant also exist in the literature~\citep{Xue, Begue, Canales2018, prd}.

One common feature of these theories is that, because of geometrical constraints (Bianchi identity), they usually require additional terms that influence the dynamics of the expansion \citep{Barrow98}. For this reason, it is suggested by~\citet{barrow2} that a Newtonian approach may be more suitable for considering cosmologies with variable $G$, in order to fully focus on the influence of the varying 
gravitational constant on the expansion dynamics. 
Newtonian cosmologies with a time-dependent $G$ have previously been theoretically investigated~\citep{barrow,Barrow1998} and, in spirit, could be considered similar to MOND theories~\citep{milgrom}, which propose modifications to Newtonian dynamics in order to explain galaxy observations without dark matter. Various other aspects of Newtonian $G(t)$ dynamics have also been studied in the literature~\citep{vinti, duval}. Following these theoretical studies, in this paper we adopt a modified Newtonian approach 
and we confront this framework with the observations. The goal of this analysis is to investigate the effect of a varying $G$ on the expansion dynamics, and see whether a modified Newtonian model with a varying $G$ can fit the low-redshift probes in a comparable way to $\Lambda$CDM without a cosmological constant or any kind of dark energy.

In this work, we focus on the low-redshift range, where a cosmological constant is relevant. We compare the predictions of our models to low-redshift probes: SNIa, baryon acoustic oscillations (BAO), and Hubble parameter, $H(z)$, measurements. 
In some models, we also allow for a redshift dependence of SNIa intrinsic luminosity in order to account for the impact of a varying $G$ on SNIa intrinsic luminosity
(see~\cite{isaac2018} and references therein for previous analyses considering this redshift dependence). On the other hand, we do not consider any data from small scales so the present discussion is only relevant for cosmological scales. As discussed earlier in relation to JBD theories, conforming to small-scale observations is a general problem for theories with a varying gravitational constant. This is generally overcome by various screening mechanisms in the literature, and similarly we will assume for the remainder of this article that the gravitational interactions of the cosmological scales are independent from the small scales.

This paper is organized as follows: in Section\,\ref{sec2} we present the derivation of our cosmological models for varying $G$ functions using a Newtonian approach. In Section\,\ref{sec3}, we describe the different cosmological probes and data sets used in this analysis, as well as the methodology adopted to fit the models to the data. The main results of this work are presented in Section\,\ref{sec4}, and we conclude in Section\,\ref{sec5}.

\section{Construction of the models}\label{sec2}

\subsection{Building a Newtonian cosmology}

The standard description of cosmological models is based on the Robertson--Walker (RW) metric, which follows from the assumption of the universe being spatially homogeneous. 
We should note that, choosing a metric does not imply general relativity, as the metric only specifies the geometry of the space. A theory of gravitation is still needed in order to obtain the dynamics of this space, which can be done with Newtonian arguments for cosmological purposes. 
The dynamics should allow the calculation of the scale factor $a(t)$ entering the RW metric. In order to derive the equation satisfied by $a(t)$, we have to rely on Birkoff--Jebsen's theorem\footnote{This famous theorem has actually been established first by the Norwegian physicist, J.T. Jebsen~\citep{Johansen2006}.}. Isolating a comoving sphere of radius $R(t)= R_0 a(t)$, we can write the mechanical energy of a test particle with mass $m$, at rest on the surface of the sphere:
\begin{equation}\label{eq0}
\frac{1}{2}m \dot{R}^2-\frac{GM}{R}m=\frac{1}{2}\alpha m\,,
\end{equation}
where the dot represents the derivative with respect to the time coordinate, the quantity $\alpha $ is a constant, and  the mass $M=4\upi R^3\rho/3$ is the integrated energy density.  Because the size $R_0$ can be chosen to be as small as one wishes, implying weak field and small velocities, this dynamics should be exactly the one obtained in general relativity~\citep{Mukhanov}. One can thereby obtain the first Friedman--Lema\^{\i}tre equation:
\begin{equation}\label{eq0b}
\left(\frac{\dot{R}}{R}\right)^2=\frac{8\uppi G \rho}{3}+\frac{\alpha}{R^2}\,.
\end{equation}

Moreover, by using the energy conservation $\dd(\rho V)= -P \dd V$, with $P$ being the pressure, one can obtain the second Friedman--Lema\^{\i}tre equation:
\begin{equation}\label{eq0c}
\frac{\ddot{R}}{R}= -\frac{4\uppi G}{3}(\rho +3 P)\,.
\end{equation}

The same equations hold for $a(t)$, identical to those obtained with general relativity. 

\subsection{What if $G$ varies with time?}

If $G$ varies  with time, energy conservation, equation\,(\ref{eq0}), cannot be assumed anymore because the gravitational force is no more conservative. In order to derive the evolution of the expansion factor, we are therefore left to apply  Newton's second law to a uniform, dust sphere of radius $R$ and mass $M$:
\begin{equation}\label{eq1}
-\frac{GM}{R^2}=\ddot{R}\,.
\end{equation}
where we used the Gauss theorem and the spherical symmetry hypothesis.
As curvature is not constrained in the Newtonian approach, we assume a flat geometry. We can now integrate equation\,(\ref{eq1}) over $R$ since it is a force equation, which gives energy when integrated over distance:
%-------------------------------------------
\begin{equation}\label{eq1_rhs}
-\int \frac{GM}{R^2}dR= \int \ddot{R}dR \,.
\end{equation}

Note that, we took Newton's gravitational constant to depend on time, $G = G(t)$, but since $R$ also is a function of time, we can integrate the above expression by a change of variables.
\begin{equation}\label{eq2}
M\left[ \frac{G}{R}  - \int \frac{dG}{R}  \right] = \frac{1}{2}\dot{R}^2 - C\,,
\end{equation}
where $C$ is an integration constant. This term looks like the spatial curvature term in general relativity~\citep{zolnierowski}, but, as noted before, we are working on a flat geometry.
In the present case, what the constant $C$ represents is the total mechanical energy in the universe at an instant. The choice of this instant is arbitrary, since this constant will ultimately be determined by the data after being written in terms of other variables.

We can then obtain from the previous relation the modified Friedmann--Lema\^itre equation, 
\begin{equation}\label{eq3}
H^2 \equiv \frac{\dot{R}^2}{R^2} = \frac{8 \uppi}{3} \rho \left[ G  - R \int  \frac{dG}{R}  \right] + \frac{2C}{R^2} \,.
\end{equation}

For simplicity, we can redefine the term inside the parentheses as another gravitational parameter, $G^*$, 
\begin{equation}\label{eq8}
G^*\equiv G - R \int  \frac{dG}{R}\,.
\end{equation}

With this definition, we essentially have two different gravitational parameters~: one, $G$, in equation~\ref{eq1} and another, $G^*$, in equation~\ref{eq3}. The equation for  the scale factor $a(t)=R(t)/R_0$, can then be derived:
\begin{equation}\label{2friedmann}
\frac{\ddot{a}}{a}=-\frac{4\uppi}{3}G\rho\quad,\quad H^2=\frac{8\uppi}{3}G^*\rho+\frac{C'}{a^2}\,,
\end{equation}
with $C'=2C / R_0^2$. These are the equivalent of the Friedmann--Lema\^itre equations. It is also convenient to define a critical energy density
\begin{equation}
  \rho_{\text{c}} \equiv \frac{3H_0^2}{ 8 \uppi  G^*_0}\,,  
\end{equation}
with $G^*_0=G^*(z=0)$. It is interesting to note that, this second gravitational parameter $G^*$ is the one most directly relevant to the cosmological discussions, since it appears in the expansion equation. As this equation is the one used in the cosmological tests, $G^*$ parameter is actually the gravitational constant that would be studied in cosmological investigations of $G$ (for instance, in \cite{zahn, galli, bai, wang}). In this sense, our modified Newtonian approach is similar to this sort of studies when the parameter $G^*$ is primarily considered.

We also introduce a density parameter $\Omega= \rho_0/\rho_{\text{c}}$. Because of mass conservation, we have $R^3\rho=\text{cst.}$, or $\rho=\rho_0/a^3=\rho_0(1+z)^3$, with $z$ being the redshift. Replacing these in equation\,(\ref{eq3}) we obtain the final form of the modified Friedmann--Lema\^itre equation used in this work
\begin{equation}
\frac{H^2(z)}{H_0^2} =  \frac{G^*(z)}{G^*_0}\Omega (1+z)^{3}  + (1-\Omega)(1+z)^{2}\,.
\label{eq:friedmann}
\end{equation}

The standard model is recovered if $G(z)$ is constant and $\Omega_\Lambda$ is added to this equation. We now consider two specific models for $G(z)$ that are expressed, for convenience, as a function of the scale factor, $a$,
\begin{equation}
(i)\ \ G(a)=G_\infty\left(1+\frac{a}{\tilde{a}} \right)\exp\left(-\frac{a}{\tilde{a}}\right)\,,
\label{eq:expo}
\end{equation}
\begin{equation}
(ii)\ \ G(a)=G_0(1 + \alpha_1(1-a)+ \alpha_2(1-a)^2+\ldots)\,,
\label{eq:power}
\end{equation}
where $G_\infty$ and $G_0$ are constants representing the values of these functions, respectively, at the early universe ($z \to\infty $) and at the present ($z=0 $).

We use the first of these equations to test two different cosmological scenarios. One scenario found in the literature is a non-accelerating model, such as the $R_h=ct$ cosmology~\citep{Melia2007,Melia2012} (see also \cite{Tutusaus2016,Moncy2019} and references therein for analyses with this kind of models). We obtain this by taking $G$ as an exponential with a positive $\tilde{a}$. Another option is a scenario where the cosmic acceleration is driven by the varying gravitational strength. Here, the same exponential equation has a negative $\tilde{a}$, which makes $G$ negative for low-redshift region (late-time), while $G^*$ remains positive.
Additionally, with the second parameterisation $(ii)$, we try a power series expansion around $a=1$ in order to look into a more general case. Unlike the exponential models, the power series models have more than one parameters, added one by one until the new parameters become redundant.

\section{Data and methodology}\label{sec3}
In this section we describe the cosmological probes and the methodology used in this work. Starting with the former, we consider late-time cosmological probes to constrain the parameters of our cosmological models, namely SNIa, BAO, and direct measurements of the Hubble parameter, $H(z)$, from cosmic chronometers.

\subsection{Type Ia supernovae}\label{sec31}
For the treatment of SNIa data we use the measurements and covariance matrix provided by the joint light-curve analysis of~\citet{betoule}. 
We obtain the observed distance modulus using the standardization method used
by the authors,
\begin{equation}
\mu_{\text{obs}}=m-M+\alpha X-\beta C\,.
\end{equation}

In this equation, $m$, $X$, and $C$ are the observed magnitude in the \emph{B}-band rest frame, and the stretch  
and colour standardization parameters for the different SNIa, respectively. These have been obtained in their analysis from the SNIa light curves and are provided in the public data set. The remaining parameters, $\alpha$, $\beta$, and $M$ are nuisance parameters, common to all SNIa, that need to be determined together with the cosmological parameters from the fit to the data. The latter is the absolute magnitude in the \emph{B}-band rest frame and, depending on the stellar mass of the host galaxy, it is given by an additional nuisance parameter $\Delta M$,
\begin{equation}
    M=
    \begin{cases}
      M', & \text{if}\ M_{\text{stellar}}<10^{10} M_\odot \\
      M'+\Delta M, & \text{otherwise}\,,
    \end{cases}
  \end{equation}
where $M_{\text{stellar}}$ is the stellar mass of the host galaxy.

There are various discussions in the literature about the effects of a varying $G$ on the intrinsic luminosity of SNIa and its dependence on redshift
~\citep{gaztanaga,mould2014,wrightli,Kazantzidis,Sakstein2019}. In this work, we follow~\citet{isaac2018} and consider an additional phenomenological standardization term to account for this redshift-dependence of the SNIa intrinsic luminosity. In this way, we can write the final expression used in this analysis for the observed distance modulus as
\begin{equation}
    \mu_{\rm obs}=m-M+\alpha X-\beta C- \epsilon z^{\delta}\,,
\end{equation}
where $\epsilon$ and $\delta$ are nuisance parameters to be determined from the fit to observations. We limit
$\delta$ to be positive and non-zero to avoid the degeneracy between $M$ and this extra term.

We compare the observed distance modulus to the predictions of the corresponding cosmological model using the expression
\begin{equation}
\mu=5\,\text{log}_{10}\left(d_L H_0\right)\,, 
\end{equation}
where the luminosity distance $d_L$ is given by
\begin{equation}
d_L=(1+z)d_M=c(1+z)\int_0^z \frac{dz'}{H(z')}\,,
\end{equation}
and $d_M$ is the comoving distance for an expanding flat space.

\subsection{Baryon acoustic oscillations}\label{sec32}
BAO in the early Universe create scales that are visible in the density distribution of galaxies. We can perform isotropic and anisotropic measurements of these scales, and relate them
back to cosmological quantities in order to constrain the parameters of a model. Isotropic measurements are sensitive to
the quantity $D_V/r_d$, with $r_d$ being the length of the standard ruler and $D_V$ given by
\begin{equation}
D_V(z)=\left( d_M^2(z) \frac{cz}{H(z)} \right)^{1/3}\,.
\end{equation}

Anisotropic measurements are sensitive to two different quantities, depending on whether the measurements are in the transverse direction:
\begin{equation}
\theta=\frac{r_d}{d_M}\,,
\end{equation}
or in the radial one:
\begin{equation}
\delta z_s=\frac{r_dH(z)}{c}\,.
\end{equation}

Let us mention that, while $r_d$ is given by the comoving sound horizon at the end of the baryon drag epoch in the concordance model, it might have a different value for models differing from $\Lambda$CDM~\citep{Verde}. Therefore, in order to be as general as possible, and not delve into the physics of the early universe, we consider $r_d$ to be a free parameter in our analysis and let the data choose its preferred value.

In this analysis, we consider the measurements from 6dFGS~\citep{beutler} at $z=0.106$, SDSS-MGS~\citep{Ross} at $z=0.15$, BOSS DR12~\citep{alam} at $z=0.38,\,0.51,\,0.61$, and eBOSS DR14~\citep{gilmarin} at $z=1.19,\,1.50,\,1.83$, as well as the Ly $\alpha$ autocorrelation function~\citep{bautista} and Ly $\alpha$-quasar cross-correlation~\citep{dumas} at $z=2.4$. We take into account the covariances for the BOSS and eBOSS measurements, we consider a correlation coefficient of $-0.38$ for the Ly $\alpha$ forest measurements, and we assume measurements of different surveys to be uncorrelated.

Due to the non-Gaussianity of the BAO observable likelihoods far from the peak, we replace the standard $\Delta \chi^2_G=-2\ln L_G$ for a Gaussian likelihood~\citep{Bassett}  by
\begin{equation}
\Delta \chi^2 = \frac{\Delta \chi^2_G}{\sqrt{1+\Delta \chi^4_G\left(\frac{S}{N}\right)^{-4}}}\,,
\end{equation}
where $S/N$ stands for the detection significance, in units of $\sigma$, of the BAO feature. We consider a detection significance of $2.4\sigma$ for 6dFGS, $2\sigma$ for SDSS-MGS, $9\sigma$ for BOSS DR12, $4\sigma$ for eBOSS DR14, and $5\sigma$ for the Ly $\alpha$ forest. Notice that some of these values are slightly lower than the ones quoted by the different collaborations in order to follow a conservative approach, and in case the likelihood becomes non-Gaussian at these high confidence levels.

\subsection{Direct \texorpdfstring{$H(z)$}{H(z)} measurements}\label{sec33}
Direct measurements of the Hubble parameter as a function of redshift can be obtained with a method called cosmic chronometers~\citep{jimenez}. This method employs observations of passive galaxies to determine their relative ages and redshifts. Since $H(z)=-(\text{d}z/\text{d}t)/(1+z)$, this method provides us with information about the recent expansion history of the Universe, independently from cosmology.

In our calculations we use $H(z)$ measurements ranging from $z=0.07$ to $z=1.965$~\citep{simon,stern,Moresco12,zhang,moresco15,moresco16}. We do not include the measurements obtained from BAO observations to avoid double counting.

\subsection{Determination of the parameter constraints}
In this work, we follow a frequentist approach and minimize the standard $\chi^2$ function to obtain the best-fitting values for the parameters of our cosmological models
\begin{equation}
\chi^2=(r_\text{model}-r_\text{data})^T C^{-1}(r_\text{model}-r_\text{data})\,,
\end{equation}
where $r_{\rm model}$ and $r_{\rm data}$ are the vectors that include the model predictions and the observed values at each redshift, respectively, and $C$ is the covariance matrix of the data. We assume SNIa, BAO, and $H(z)$ measurements to be statistically independent, therefore their $\chi^2$ values can be added together. We minimize this function using the \texttt{IMINUIT} library\,\footnote{\url{https://pypi.org/project/iminuit/}} of \texttt{PYTHON}. It is an implementation of \texttt{SEAL MINUIT}, which is a minimizer developed at CERN~\citep{iminuit}. As we do not conceive the models in this work to be competitors to the standard model, we only present these goodness-of-fit values and refrain from using a model comparison criteria such as AIC \citep{akaike}. 

We fit the predictions of our models for the different observables to 789 data points in total, where 740 correspond to  SNIa, 16  to BAO, and 30 to $H(z)$ measurements. For the SNIa observations, we have four nuisance parameters given by $\alpha$, $\beta$, $M'$, and $\Delta M$, which are described in Section\,\ref{sec31}, and two extra nuisance parameters, $\epsilon$ and $\delta$, when we account for the possibility of SNIa intrinsic luminosity evolution. Specific to our models with evolving $G$, in addition to the cosmological parameter $\Omega$, we have one extra nuisance parameter, $\tilde{a}$, for the exponential model and between one and two parameters, $\alpha_1$--$\alpha_2$, in the power series model. When considering BAO measurements we add two cosmological parameters to the analysis, $H_0$ and $r_d$, as described in Section\,\ref{sec32}. Notice that these two parameters are completely degenerate when only SNIa and BAO data are taken into account. The introduction of measurements on $H(z)$ breaks this degeneracy and allows us to constrain both parameters at the same time.

\section{Results and discussion}\label{sec4}
One interesting feature of our calculations is the appearance of two different cosmological constants, $G$ and $G^*$, in the two Friedmann--Lema{\^i}tre equations, equation\,(\ref{2friedmann}). However, this is not necessarily specific to our models. While we essentially modified the first equation of equation\,(\ref{2friedmann}), and used this to derive the other, in the literature the phenomenology often adopted is the other way around, modifying the gravitational constant directly in the second equation, i.e. $G^*$ is taken as the gravitational constant (as in~\cite{zahn, galli, bai, wang}, for instance). Then, in these models, if $G^*$ increases sufficiently rapidly in time, it is possible to obtain an accelerated expansion in the universe. This surprising result becomes clear within our Newtonian approach. Indeed, taking the time derivative of equation\,(\ref{eq:friedmann}), we obtain
\begin{equation}
    \ddot{a}= \frac{H_0^2 \Omega}{2 G^*_0}\frac{d}{da}(a^{-1}G^*)\,.
\end{equation}

Therefore, if $a^{-1}G^*$ is increasing at the late stages, $\ddot{a}$ will be positive, implying an accelerated expansion. Also, since
\begin{equation}
    G=-\frac{d}{da}(a^{-1}G^*)a^{2}\,,
\end{equation}
$G$ will become negative for the same condition. 
This means that, if the variation of the gravitational constant is the only factor causing acceleration, even when $G^*$ is positive in the cosmological equation, $G$ in the force equation can still be negative.

We present the $\chi^2$ values for the tested models in Table~\ref{table1} along with the flat $\Lambda$CDM results for comparison. The best-fitting values of the cosmologically relevant parameters are given in Table~\ref{tab:table2}. We do not include the nuisance parameters for SNIa in these tables, since they do not change appreciably between the models and their small variations do not significantly contribute to the analysis. In Table~\ref{table1} we can see that most of the considered models are able to achieve somewhat lower $\chi^2$ values compared to the flat $\Lambda$CDM case. The exceptions, model 4 (the exponential with positive parameter and SNIa luminosity evolution), and model 5 (one-parameter power series without SNIa luminosity evolution) also have low $\chi^2$ values, comparable to the standard model.
We saw that, when considering the power series models adding higher order terms than the second does not improve the fit, so their results are not included in these tables. On the other hand, we rule out the exponential model with positive parameter without SNIa luminosity evolution (model 3) since this model has a high $\chi^2$, and we do not discuss it further.

\begin{table*}
  \centering
    \caption{$\chi^2$ values over degrees of freedom for the considered models. PS refers to power series models. Exponential (pos.) refers to the exponential with positive $\tilde{a}$ and exponential (neg.) refers to the exponential with negative $\tilde{a}$. 
  }\label{table1}
    \begin{tabular}{|c | c c c c c  c c|} % <-- Alignments: 1st column left, 2nd middle and 3rd right, with vertical lines in between
    \hline
     \footnotesize Model &\footnotesize $G(z)$ model & \footnotesize SNIa lum. evo. &\footnotesize $\chi^2$ & Degrees of freedom& $\chi^2$/Degrees of freedom &\footnotesize $G_0$ &\footnotesize $G^*_0$\\
      \hline
\footnotesize0. & \footnotesize$\Lambda$CDM  & \footnotesize NO & \footnotesize712.79 & 782 & 0.91 & \footnotesize Positive & \footnotesize Positive \\
\footnotesize1. &\footnotesize Exponential (neg.) &\footnotesize NO &\footnotesize 711.41 & 781&  
0.91 & \footnotesize Negative &\footnotesize Positive\\
\footnotesize 2. &\footnotesize Exponential (neg.) &\footnotesize YES &\footnotesize 709.68 & 779& 0.91 &\footnotesize Negative &\footnotesize Positive\\
\footnotesize 3. &\footnotesize Exponential (pos.) &\footnotesize NO &\footnotesize 755.55 & 781& 0.97 &\footnotesize Positive &\footnotesize Positive\\
\footnotesize 4. &\footnotesize Exponential (pos.) &\footnotesize YES &\footnotesize 714.44 & 779& 0.92& \footnotesize Positive & \footnotesize Positive\\
\footnotesize 5. &\footnotesize PS (1-parameter) &\footnotesize NO &\footnotesize 716.75 &781& 0.92 &\footnotesize Negative &\footnotesize Negative\\
\footnotesize 6. &\footnotesize PS (1-parameter)&\footnotesize YES &\footnotesize 710.34&779& 0.91&\footnotesize Negative &\footnotesize Negative\\
\footnotesize 7. &\footnotesize PS (2-parameter) &\footnotesize NO &\footnotesize 709.75&780& 0.91&\footnotesize Negative &\footnotesize Positive\\
\footnotesize 8. &\footnotesize PS (2-parameter)&\footnotesize YES &\footnotesize 707.62&778& 0.91&\footnotesize Negative &\footnotesize Positive\\

\hline
    \end{tabular}
\end{table*}

\begin{table*}
%    \begin{subtable}
%   \centering
             \caption{Best-fitting values of the cosmological parameters of the considered $G(z)$ models.
             }
     \label{tab:table2}
     
        \begin{tabular}{|c|c c c c c c c c|} % <-- Alignments: 1st column left, 2nd middle and 3rd right, with vertical lines in between
    \hline
      \footnotesize Model & \footnotesize $\tilde{a}$ & \footnotesize $\alpha_1$ & \footnotesize $\alpha_2$ & \footnotesize{$\Omega$ $^a$}& \footnotesize{$\epsilon$}& \footnotesize{$\delta$}& \footnotesize{$H_0$ $^b$}& \footnotesize{$r_d$ $^c$}\\
     % $\alpha$ & $\beta$ & $\gamma$ \\
      \hline
    \footnotesize 0. & \footnotesize - & -  & - & \footnotesize  \(0.294  \pm 0.016\) & - & - &\footnotesize  \( 69.3  \pm 1.8 \)&\footnotesize  \( 146   \pm 4  \) \\      
%    \hline
    
     \footnotesize 1. & \footnotesize \( -0.59  \pm 0.02 \) & - & - & \footnotesize  \(1.50  \pm 0.18\) & - & - &\footnotesize  \( 68.6  \pm 1.9 \)&\footnotesize  \( 146   \pm 4  \) \\  
     \footnotesize 2.& \footnotesize \(-0.61  \pm 0.03 \) & - & - &
     \footnotesize  \(1.20  \pm 0.31\) & \footnotesize \( 0.12 \pm 0.14\)&\footnotesize  \( 0.4 \pm 1.8\)&\footnotesize  \( 66.4  \pm 2.5\)&\footnotesize  \( 147  \pm 4 \) \\
     \footnotesize 3.& \footnotesize \(0.07   \pm 0.04 \) & - & - &
      \footnotesize  \( (3.6 \pm 18) \times 10^{-6} \) & \footnotesize -&\footnotesize -&\footnotesize  \( 62.1 \pm 1.2 \)&\footnotesize  \( 147 \pm 3   \) \\

     \footnotesize 4.& \footnotesize \(0.10 \pm 0.06 \) & - & - &
     \footnotesize  \((5.6 \pm 23) \times 10^{-5} \) & \footnotesize \(0.32  \pm 0.062\)&\footnotesize  \( 0.42   \pm 0.20\)&\footnotesize  \( 62.0  \pm 1.5\)&\footnotesize  \( 148   \pm 4  \) \\
\hline

   \footnotesize 5.& - & \footnotesize \(-2.16  \pm 0.07 \) & - &\footnotesize \( -0.68  \pm  0.09\)&-&-&\footnotesize \(67.9  \pm 1.9  \)&\footnotesize \(146   \pm  4 \)\\ 
      
  \footnotesize 6.& - &\footnotesize \(-2.31  \pm 0.20 \) & - &\footnotesize \(-0.42 \pm 0.17\)&\footnotesize \( 0.26  \pm 0.35 \)&\footnotesize \(0.20  \pm 0.47 \)&\footnotesize \(65.3   \pm 2.3 \)&\footnotesize \(147  \pm 4 \)\\
  
   \footnotesize 7. & - & \footnotesize \( -4.21   \pm 0.65 \) &\footnotesize \(3.9 \pm 1.2\) & \footnotesize \(8.1 \pm 4.5\)& -& -&\footnotesize \(68.5  \pm 1.9 \)&\footnotesize \(147   \pm 4 \)\\ 
   
   \footnotesize 8. & -  &\footnotesize \( -4.45  \pm 0.68 \) &\footnotesize \(4.3  \pm 1.3\) & \footnotesize \(8.8 \pm 4.5\)& \footnotesize \( 0.08  \pm 0.05\)& \footnotesize \(2.5 \pm 2.4\)&\footnotesize \(67.6  \pm 2.0 \)&\footnotesize \(148  \pm 4 \)\\

\hline
    \end{tabular}
%     \end{subtable}
     \newline
     \footnotesize{$^a$ $\Omega$ is negative if $G^*_0$ is negative, $^b$ $H_0$ has the units (km s$^{-1}$ \text{Mpc}$^{-1}$), $^c$ $r_d$ has the units (Mpc)}.

\end{table*}

\begin{figure}
\centering
  \includegraphics[width=1.0\linewidth]{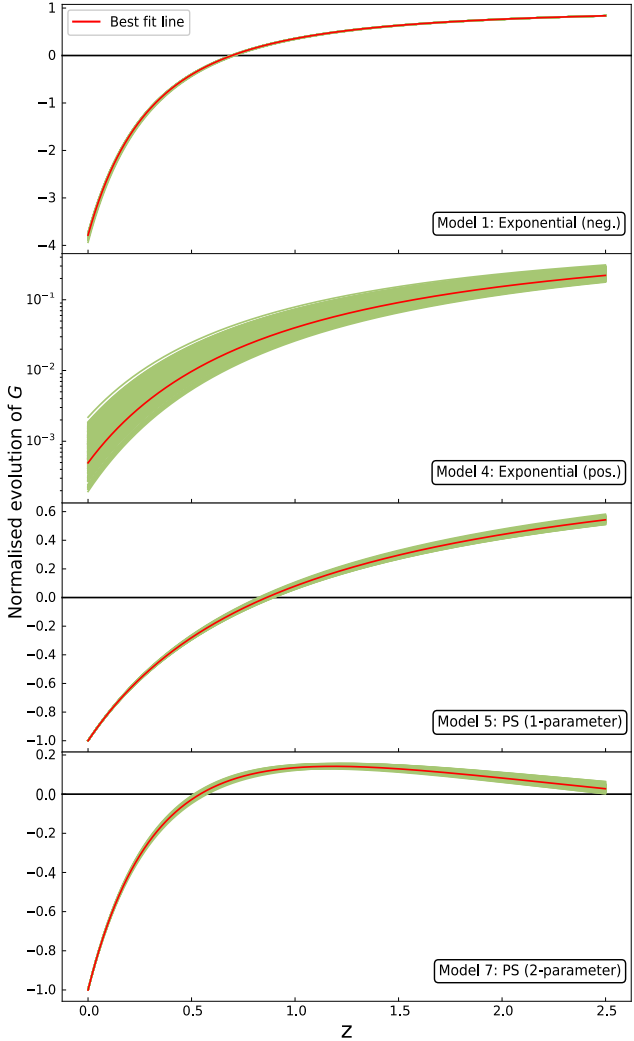}

\caption{Variation of $G$ versus redshift for models 1, 4, 5, and 7 from Table\,\ref{table1} from top to bottom. The red lines represent $-G(z)/G_0$ (for power series) or $G(z)/G_{\infty}$ (for exponential) using the best-fitting values for the different parameters, while the green bands encapsulate the reconstructions with $\Delta \chi^2<1$ (see the text for details).
}\label{fig1}
\end{figure}

\begin{figure}
\centering
  \includegraphics[width=1.0\linewidth]{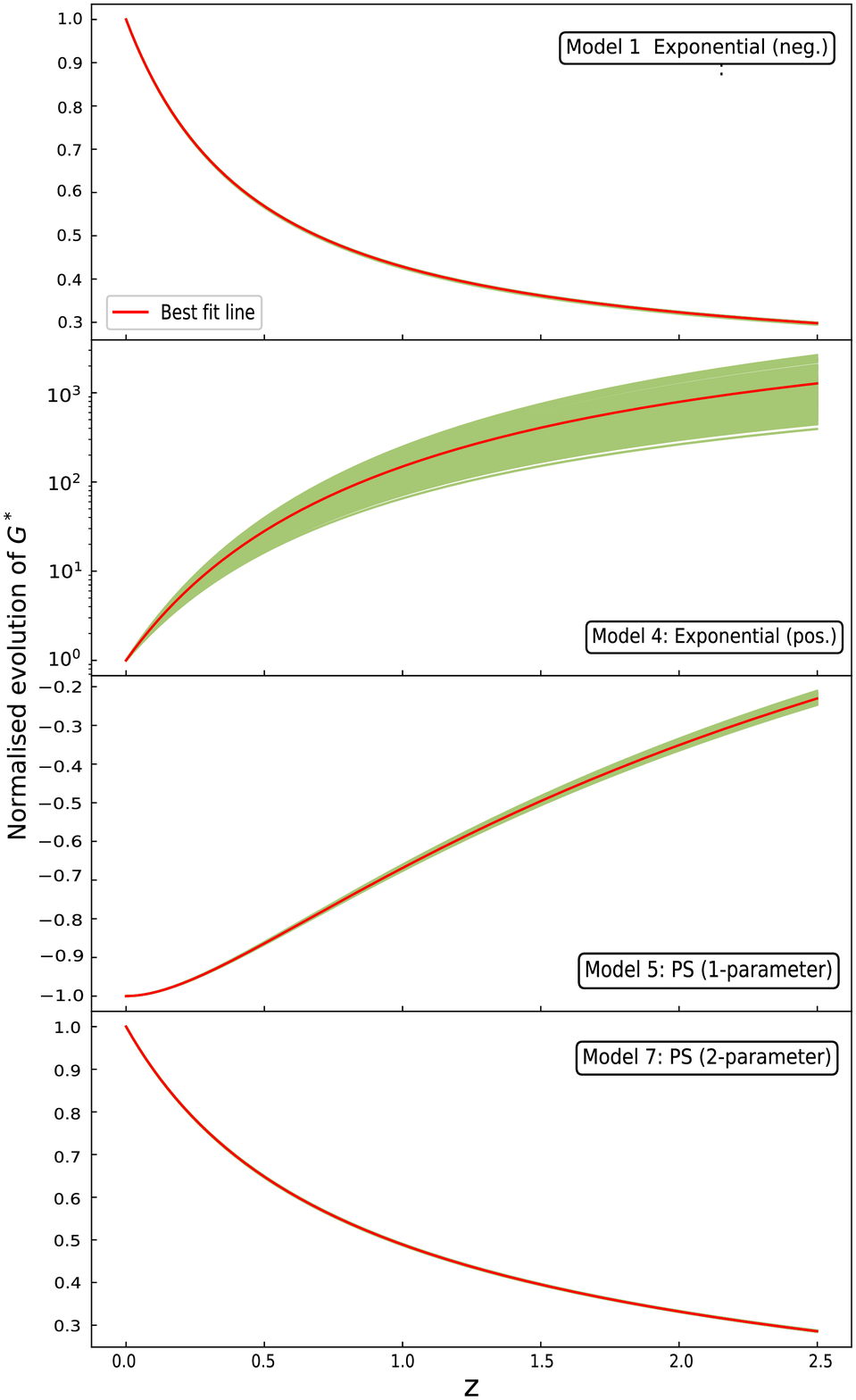}
  
\caption{Variation of $G^*$ versus redshift for models 1, 4, 5, and 7 from Table\,\ref{table1} from top to bottom. The red lines represent $G^*(z)/|G^*_0|$ using the best-fitting values for the different parameters, while the green bands encapsulate the reconstructions with $\Delta \chi^2<1$ (see the text for details).
}\label{fig2}
\end{figure}

\begin{figure}
 \centering
  \includegraphics[width=1.0\linewidth]{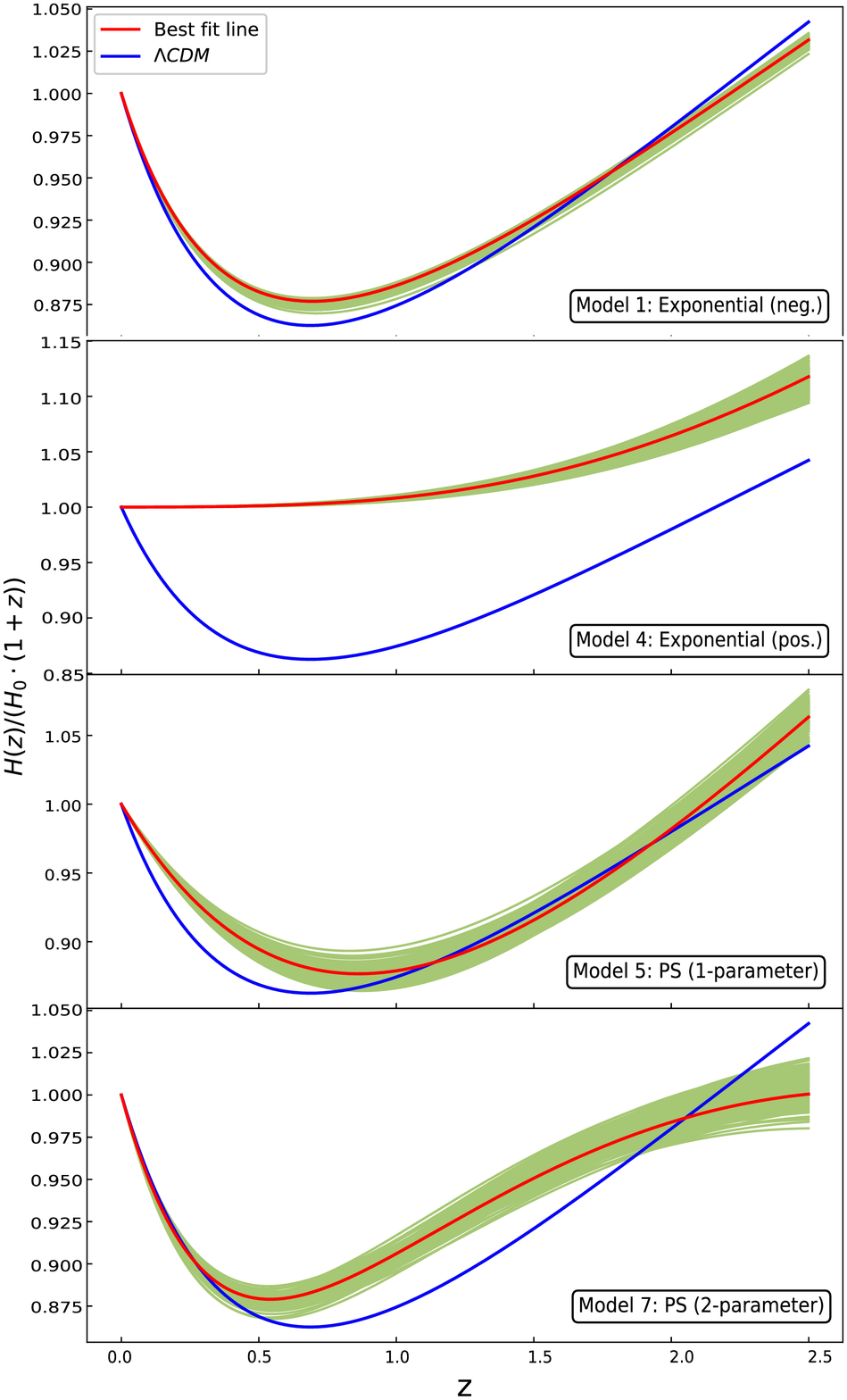}
  \caption{Dimensionless expansion rate of the universe, as a function of redshift, over $(1+z)$ for models 1, 4, 5, and 7 from Table\,\ref{table1} from top to bottom. The red lines represent the curve obtained with the best-fitting values of the parameters, while the green bands show the reconstructions within $\Delta \chi^2<1$ (see the text for details). The blue lines stand for the $\Lambda$CDM model. 
  }\label{fig3}
\end{figure}

When we compare the $\chi^2$ values of the equivalent models with and without SNIa evolution in Table\,\ref{table1}, we can see a slight improvement in each model when SNIa luminosity is allowed to vary as a function of redshift. However, since the values of the parameter $\epsilon$ in Table~\ref{tab:table2} are compatible with zero within one sigma for models 2, 6, and 8, we can conclude that most of our models, with the exception of the positive exponential (model 4), do not necessarily require SNIa luminosity evolution in order to fit the data adequately.
Therefore, in Figues~\ref{fig1},~\ref{fig2} and~\ref{fig3}, we only present models 1, 4, 5, and 7. Similarly, when we include SNIa luminosity evolution to the $\Lambda$CDM model for comparison $\chi^2$ does not improve, with $\epsilon$ being consistent with zero. Therefore, we do not present these results.

Turning to Table~\ref{tab:table2}, we see that the results of $H_0$ and $r_d$ are mostly comparable between different models. We note that the best-fitting values of $H_0$ are somewhat smaller for the models that allow for SNIa luminosity evolution. However, given the error bars for $H_0$, this difference is not that significant. 
Consequently, models 2, 6, and 8 are still consistent with their counterparts without SNIa luminosity evolution (models 1, 5, and 7). This can be expected since the SNIa luminosity evolution parameters are also compatible with zero.
For the exponential with positive parameter (model 4), this situation is quite different and we can confirm the effect of SNIa evolution on $H_0$ by noting that this model predicts $H_0=62.0 \pm 1.46$ km s$^{-1}$ \text{Mpc}$^{-1}$, which is consistent with earlier results by~\cite{isaac2018}. It is evident, therefore, that adding the luminosity evolution to SNIa tends to decrease $H_0$, meaning that the speed of expansion implied by the cosmological data becomes lower.

There is a significant divergence in the literature between the $H_0$ measurements given by different sources and methods (for further discussion see~\citet{2019NatAs...3..891V}). For instance, the Cepheid calibrated local observations provide $H_0=74.22 \pm 1.82$ km s$^{-1}$ \text{Mpc}$^{-1}$~\citep{Riess_2019}, while, assuming $\Lambda$CDM, Planck collaboration obtains the value $H_0 = 67.4 \pm 0.5$ km s$^{-1}$ \text{Mpc}$^{-1}$~\citep{planck}. Our calculations with zero or negligible SNIa evolution agree quite well with the Planck results. Therefore, we are able to replicate a similar behaviour as the standard model without having $\Lambda$.

Comparing the results obtained for $r_d$ with model independent estimates, we see that, as expected, the values of $r_d$ are not affected by the variation in SNIa luminosities. The values in Tables~\ref{tab:table2} for all the models are consistent with $r_d=145.61^{+2.82}_{-7.12}$ Mpc, given by~\cite{haridasuJCAP}, as well as $r_d=147.4 \pm 0.7$ Mpc, obtained from CMB measurements by~\cite{Verde2}. Since we do not consider the early universe physics in this work, this consistency might be spurious. Indeed, if $G$ has a different value during the radiation epoch, the CMB physics (as well as the Big Bang Nucleosynthesis) would be affected (as discussed in~\citet{Uzan2011}), but these effects of $G$ depend on the gravitational theory. A more detailed discussion of this is outside the scope of this work. However, we might say that these obtained $r_d$ values imply a similar behaviour to the standard cosmology on the background expansion level. 

As discussed previously in Section~\ref{sec2}, the actual value of $G$ is proportional to a coefficient ($G_0$ or $G_\infty$) in
equations\,(\ref{eq:expo}-\ref{eq:power}). However, these coefficients cancel out in equation\,(\ref{eq:friedmann}), used to fit the data. In other words, the cosmological probes used in this analysis are not sensitive to the values of $G_0$ or $G_{\infty}$. For this reason, it is necessary to use additional arguments in order to determine the exact value of $G$. However, without making any additional assumptions, we can obtain the sign of these coefficients by requiring that the energy density of matter is positive. Since we have
\begin{equation}
  \Omega=\rho \frac{ 8 \uppi  G^*_0}{3H_0^2}\,, 
\end{equation}
if $\rho$ is positive, the signs of $\Omega$ and $G^*_0$ should be the same.
We compute $G^*_0/G_0$ for the power series models and $G^*_0/G_\infty$ for the exponential models (from equations \ref{eq8}, \ref{eq:expo}--\ref{eq:power}) using the best-fitting values for $\alpha_1$, $\alpha_2$, and $\tilde{a}$ (in Table~\ref{tab:table2}). With these and the corresponding $\Omega$ values, we deduce the signs of the respective $G(a)$ functions for each model. The resulting signs of $G$ and $G^*$ are given with respect to their values at the present epoch ($G_0$ and $G^*_0$) in Table~\ref{table1}. This table shows that, for $G_0$, with the exception of the positive exponential, all considered models have negative $G_0$ today. This, of course, contradicts laboratory and Solar System experiments. However, for the purposes of this discussion the only relevant point is that some sort of mechanism is indeed needed to screen large-scale gravitation from small scales in order to embrace these kind of models. On the other hand, most of the $G^*_0$ values turn out to be positive. In fact, Figure~\ref{fig2} illustrates that $G^*$ never becomes negative for neither the exponential models (models 1 and 4), nor the two-parameter power series model (model 7).

Figures\,\ref{fig1} and\,\ref{fig2} show the normalized evolution of $G$, and $G^*$, respectively, as a function of redshift. We present in the top panel the negative exponential model (model 1), the positive exponential model (model 4) in the second panel, and the power series models with one and two parameters (models 5 and 7) are shown in the third and the bottom panels, respectively. The red lines in these figures show the variation of $-G/G_0$ or $G/G_{\infty}$, and $G^*/G^*_0$ as a function of redshift using the best-fitting values obtained from the fit to the cosmological data sets used in this work, while the uncertainties 
are shown as green bands. These bands are drawn by randomly generating values for the parameters from Gaussian distributions centred at the best-fitting values and width determined by the obtained standard deviations. We then select only the $G(z)$ reconstructions whose $\Delta \chi^2$ (compared to the best-fitting $\chi^2$) is smaller than one. We present the power series models in the bottom two panels of Figure~\ref{fig1} with the normalisation $-G_0$, to emphasize that $G_0$ is found to be negative for these models. Likewise, \emph{y}-axis of Figure~\ref{fig2} is given by $G^*/|G^*_0|$ in order to emphasize the models with $G^*_0<0$.

Figure~\ref{fig3} displays the behaviour of the normalized expansion rate of the universe for the same models shown in Fig.~\ref{fig1} and~\ref{fig2} along with $\Lambda$CDM, in blue, for comparative purposes. Note that since we are normalizing the expansion rate by a factor $(1+z)$, the minima of these lines provide %Additionally, the minimum points of the y-axis in this graphs provide 
the transition redshift, $z_T$. The first panel shows the negative exponential model (model 1 in Table~\ref{table1}) to behave very similarly to the standard model, transitioning to the accelerated expansion at $z_T = 0.69 \pm 0.01$,  while the power series model with one parameter (model 5)
shows an earlier transition, at $z_T = 0.86 \pm 0.02$. The two-parameter power series model (model 7) shows almost the exact same expansion rate as $\Lambda$CDM after $z \approx 0.35$ but it differs to a larger degree in higher redshifts than the one-parameter power series model. This model shows an even later transition to accelerated expansion, at $z = 0.54 \pm 0.01$.
The presented error bars show the standard deviations of the transition redshifts of the green $\chi^2<1$ bands in Figure~\ref{fig3}.
These results are consistent with $z_T=0.64^{+0.12}_{-0.09}$, calculated from a model independent $H(z)$ reconstruction in~\cite{haridasuJCAP}, within one or two sigma depending on the model. \cite{moresco16} give another model independent estimation of the transition redshift as $z_T=0.4 \pm 0.1$ from the five $H(z)$ measurements they provide. While our models have larger transition redshifts than this, our results are better compatible with their other result, also provided in the same paper as $z_T=0.64^{+0.11}_{-0.06}$, which takes into account other cosmic chronometer measurements. Yet another model independent estimate from SNIa, BAO, and cosmic chronometer data is given by~\cite{GomezValent2019} as $z_t=0.8 \pm 0.1$ using the Weighted Function Regression method~~\citep{GomezValent2018}, which is closer to our estimates for models 1 and 5 than the other cited results.

One interesting quality of the exponential model with positive parameter and SNIa luminosity evolution is that this model does not show late-stage acceleration, as seen by the second panel of Figure~\ref{fig3}. Since the $\chi^2$ for this model is also low (in Table~\ref{table1}), this indicates that, if SNIa intrinsic luminosity varies with cosmic time, low-redshift cosmological observations are consistent with non-accelerated dynamics, in agreement with earlier results~\citep{isaac2017}.

Moreover, we can see that there is a correlation between the acceleration of different models in Figure~\ref{fig3} and the behaviour of $G^*$ in Figure~\ref{fig2}. As discussed before, $G^*$ has to be increasing with time for the accelerated models, which corresponds to a negative $G$ to drive the acceleration. Since these are also the models without the luminosity evolution of SNIa, this indicates that the acceleration is required by the late Universe data if SNIa are assumed to have constant intrinsic luminosity.

Another thing to note about these results is that, even though we derive our equations in a Newtonian framework, the analysis in this work would still be valid under a different theory, provided that the dynamics for the late stage expansion are analogous to equation\,(\ref{eq:friedmann}). 
We can assume an underlying higher order theory that gives equation\,(\ref{eq:friedmann}) with negative $G$ in large scales while reducing to the usual general relativity in small scales, or high density regions. An example of a similar situation occurs in conformal gravity, as discussed by~\cite{mannheim}. In this case, the cosmological gravity may decouple from the gravity of the local scales, because of the differing levels of symmetry between the relevant metrics. Then, spontaneous symmetry breaking can lead to $G$ being negative in the cosmological context, while the usual gravitational interactions are preserved locally.

\section{Conclusion}\label{sec5}
In this work we have investigated modified Newtonian cosmologies with a time-varying $G$. We have seen that various interesting cosmological scenarios, including some that can be found in the literature, can be obtained using this framework. We have shown that models with a variable gravitational constant in general have two different gravitational parameters in the two equivalent Friedmann--Lema{\^i}tre equations. One of these parameters corresponds to the gravitational constant in Newton's second law, while the other appears in the first Friedmann--Lema{\^i}tre equation. Further investigating these equations, we see that in an accelerated scenario, where SNIa are assumed to be standard candles, at least one of them has to be negative without an additional component to cause acceleration.

As a result of testing these varying $G$ models, we have seen that, without a cosmological constant, they are compatible with low-redshift cosmological data. 
A model with exponentially evolving $G$ (model 4) is shown to fit the low-redshift cosmological observations quite well when taken with a redshift-dependent SNIa intrinsic luminosity. This model has a positive gravitational constant, $G$, that decays to zero at late times, leading to a non-accelerated universe. Therefore, we see that a redshift dependence of SNIa can relieve the need for cosmic acceleration. The results for this model are also consistent with earlier analysis of $R_h=ct$ cosmologies.

On the other hand, another exponentially varying model (model 1) for $G$ is shown to fit the data in a similar way to the flat $\Lambda$CDM model. This model also has an almost constant $G$ in the high-$z$ regime, which would likely allow us to match the high-redshift observations. 
This model does not need a modification in the SNIa luminosity but requires a negative cosmological $G$ value in the present to propel the acceleration.

When $G$ is expanded as a second order power series around the present epoch, we have shown that Newtonian cosmological models (models 5 to 8) can have an accelerated expansion similar to the one created by a cosmological constant with a similar $\chi^2$ value compared to the standard model. 
Even though this expansion is only accurate in the low redshifts, we see that these models are able to adequately conform to the late Universe observations.

In conclusion, we see that the modified Newtonian approach proves to be a suitable test-lab for considering various cosmological possibilities that can be found in the literature. Despite their already discussed shortcomings, these models offer interesting interpretations of the low-redshift cosmological data without requiring a dark energy fluid. Therefore, this work shows the interest of studying more general variable $G$ models.

\section*{Data Availability}

No new data are generated in support of this research. All the analysed data are obtained from the works cited in Sections\,\ref{sec31}, \ref{sec32}, and \ref{sec33} of this article.

\section*{Acknowledgements}
This work has been funded in part with support from the European Commission through Erasmus+ traineeship grant received by ETH. This publication reflects the views only of the authors, and the Commission cannot be held responsible for any use which may be made of the information contained therein.

%%%%%%%%%%%%%%%%%%%%%%%%%%%%%%%%%%%%%%%%%%%%%%%%%%



\bibliographystyle{mnras}
\bibliography{reference.bib} % if your bibtex file is called example.bib

%\begin{thebibliography}{99}
%\end{thebibliography}

%%%%%%%%%%%%%%%%%%%%%%%%%%%%%%%%%%%%%%%%%%%%%%%%%%

%%%%%%%%%%%%%%%%% APPENDICES %%%%%%%%%%%%%%%%%%%%%

%\appendix



%%%%%%%%%%%%%%%%%%%%%%%%%%%%%%%%%%%%%%%%%%%%%%%%%%

% Don't change these lines
\bsp	% typesetting comment
\label{lastpage}
\end{document}